\documentclass[%
 reprint,
 amsmath,amssymb,
 aps,
]{revtex4-1}

\usepackage{graphicx}
\usepackage{dcolumn}
\usepackage{bm}

\begin{document}

\preprint{APS/123-QED}

\title{Optical cavity for enhanced parametric four-wave mixing in rubidium}

\author{E. Brekke}%
\email{erik.brekke@snc.edu}
\author{S. Potier}
 
\affiliation{%
 Physics Department, St. Norbert College, De Pere, WI 54115
}%

\date{\today}

\begin{abstract}
We demonstrate the implementation of a ring cavity to enhance the efficiency of parametric four-wave mixing in rubidium. Using an input coupler with 95$\%$ reflectance, a finesse of 19.6$\pm$0.5 is achieved with a rubidium cell inside.  This increases the circulating intensity by a factor of 5.6$\pm$0.5, and through two-photon excitation on the $5s_{1/2}\rightarrow5d_{5/2}$ transition with a single excitation laser, up to 1.9$\pm$0.3 mW of power at 420 nm is generated,  50 times what was previously generated with this scheme.  The dependence of the output on Rb density and input power has been explored, suggesting the process may be approaching saturation.  The blue output of the cavity also shows greatly improved spatial quality, combining to make this a promising source of 420 nm light for future experiments.
\begin{description}

\item[OCIS numbers]
(190.4380) Nonlinear optics, four-wave mixing; (190.7220) Upconversion;\\ (140.4780)    Optical resonators.

\end{description}
\end{abstract}

\maketitle



\section{Introduction}

Four-wave mixing continues to be an active area of research, with applications in a number of different contexts and materials.  Currently, four-wave mixing is being pursued in applications including quantum information \cite{Camacho:09, Radnaev:10}, Rydberg States \cite{Chen:16, deMelo:14, Brekke:08},  single-photon sources \cite{Gulati:14}, and optical gyroscopes \cite{Mikhailov:14}.  It has been examined in a wide range of contexts, including hollow core fibers \cite{Londero:09}, cold atoms \cite{Srivathsan:13} and thermal vapors of rubidium and cesium \cite{Zibrov:02,Schultz:09}.

One of the main applications of four-wave mixing continues to be frequency up-conversion in rubidium, both using a two-step excitation process \cite{Meijer:06, Akulshin:09, Vernier:10, Lopez:16}, and using a single laser for two-photon excitation \cite{Brekke:13, Cao:15}.  Extensive work has already been done to better understand the linewidth and frequency characteristics of the resulting light \cite{Akulshin:14, Akulshin:12, Brekke:15}.  Recently the infrared light generated has also been examined \cite{Akulshin:14b, Sell:14}, and this process has been used to investigate the transfer of angular momentum \cite{Walker:12, Akulshin:15}.  Several improvements have been made to the two-step process, including optical pumping \cite{Akulshin:12b}, optimizing excitation frequency \cite{Vernier:10}, high input powers \cite{Sell:14}, and using a cavity for the light produced \cite{Offer:16}, leading to output powers on the order of a mW.  Meanwhile, the two-photon process has been limited to either pulsed systems or low output powers.  

In this paper we demonstrate the implementation of a ring cavity to dramatically increase the power generated at 420 nm for parametric four-wave mixing in rubidium using a single laser excitation scheme. Using a low-finesse build up cavity for the 778 nm excitation light, the circulating intensity was increased by a factor of 5.6$\pm$0.5. This led to an increase in blue power by over two orders of magnitude, with a maximum power achieved of 1.9$\pm$0.3 mW, comparable to the highest output powers achieved in the two-step process.  We have investigated the scaling of the output power with cell temperature and input intensity, as well as investigated the optimal parameters for the cavity.  This technique presents a simple and attractive method for the generation of tunable far infrared light and blue light near the $5s_{1/2}\rightarrow6p_{3/2}$ transition in rubidium.  

\begin{figure}[htbp]
\centering
\includegraphics[width=8.2 cm]{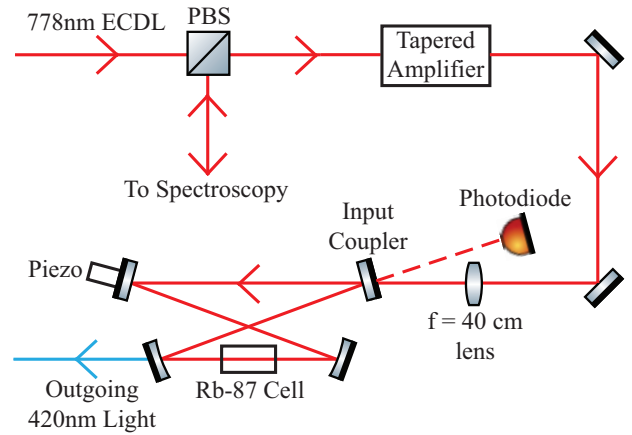}
\caption {A simplified version of the experimental setup.  A single ECDL laser on the $5s_{1/2}\rightarrow5d_{5/2}$ transition is focused through a Rb cell inside a ring cavity.}
\label{fig:ExpSetup}
\end{figure}

\section{Experimental Setup and Cavity Design}
\label{sec:setup}

\begin{figure}[htbp]
\centering
\includegraphics[width=8.2cm]{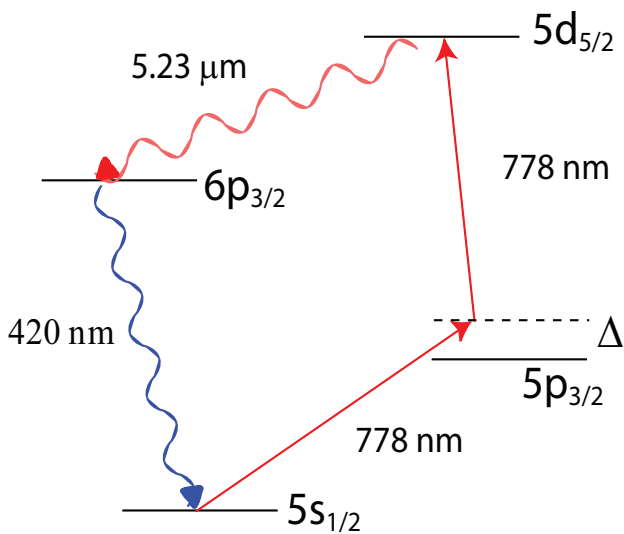}
\caption {Energy levels involved in the FWM process. Two-photon excitation is accomplished using a single 778 nm ECDL on the $5s_{1/2}\rightarrow5d_{5/2}$ transition.  The resulting process produces coherent and collimated beams at 5.23 $\mu$m and 420 nm.}
\label{fig:levels}
\end{figure}

Our experimental setup is schematically illustrated in Fig. \ref{fig:ExpSetup}. A single ECDL at 778 nm excites the two photon $5s_{1/2}\rightarrow5d_{5/2}$ transition in rubidium.  The frequency control and tapered amplifier system have been described previously \cite{Brekke:13}.  Amplified spontaneous emission and four-wave mixing in rubidium result in generated beams at 5.23 $\mu$m and 420 nm, with the relevant energy levels shown in Fig. \ref{fig:levels}. Here, we introduce a ring cavity surrounding a heated rubidium cell, to increase the circulating intensity and dramatically increase the power generated in the non-linear four-wave mixing process.

\begin{figure}[htbp]
\centering
\includegraphics[width=\linewidth]{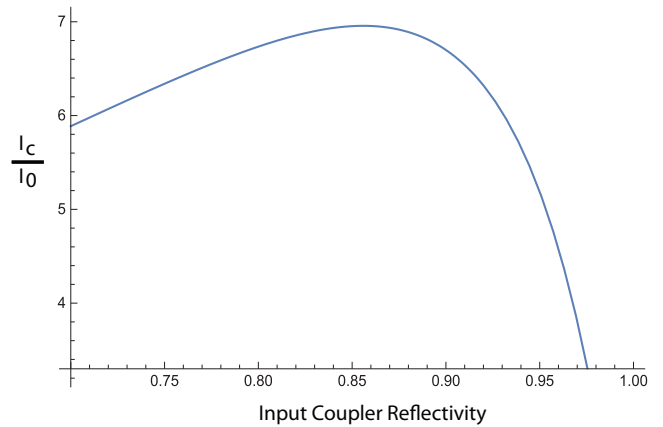}
\caption{A plot of the ratio of circulating intensity to incident intensity in the ring cavity as a function of input coupler reflectance.  This graph is for a cell transmission of 85$\%$.  }
\label{fig:intensity}
\end{figure}

The ring cavity is designed to focus the beam to a small waist inside a heated rubidium cell.  Typical Pyrex cells have transmissions in the 80-90$\%$, limiting the possible finesse of the cavity.  A 95$\%$ reflectance input coupler was used to increase the circulating intensity in the cavity, with the other three mirrors reflecting more than 99.5$\%$ at 778 nm.  The output coupler has more than 85$\%$ transmission at 420 nm.   The total circulating intensity inside the cell at maximum is given by 
\begin{equation}
\frac{I_c}{I_0}=\frac{1-R}{1-2\sqrt{RT_c}+RT_c},
\label{eqn:intensity}
\end{equation}
Where $I_c$ is the circulating intensity, $I_0$ is the incident intensity, $R$ is the reflectivity of the input coupler, $T_c$ is the transmission through the cell, with the reflectivity of the other mirrors approximated as 100$\%$. The transmission through our cell was 86$\%$, and a plot of the maximum circulating intensity vs input coupler reflectance is shown in Fig. \ref{fig:intensity}.  The current work was done with an input reflectance of 95$\%$, giving a theoretical intensity gain of  $5.2$.  Further optimization should be possible with an even lower reflectance at the optimal value of 86$\%$.

To measure the finesse and circulating power in the cavity, a photodiode was used to examine the reflectance off the input coupler, which is combined with the transmission of the circulating light, as shown in Fig. \ref{fig:ExpSetup}.  As the piezo voltage is varied, certain cavity lengths give constructive interference in the cavity.  The cavity had a finesse of 19.6$\pm$0.5, and gave a circulating power of 5.6$\pm$0.5 times the original, consistent with theoretical expectations.

The cavity is designed to have two waists, one between the plane mirrors and one between the R=10 cm spherical mirrors.  A lens with focal length 40 cm was used to couple the light into the cavity more effectively.  The cavity results in a waist of $88~\mu$m between the plane mirrors and $22 ~\mu$m  between the spherical mirrors.  With 1.5 W input power, and the circulating intensity being $5.6$  times that at $8.4$  W,  the maximum intensity in the Rb cell is  $1.1\times 10^{10} \frac{W}{m^2}$.

\section{Results and Analysis}

\begin{figure}[htbp]
\centering
\includegraphics[width=8.6 cm]{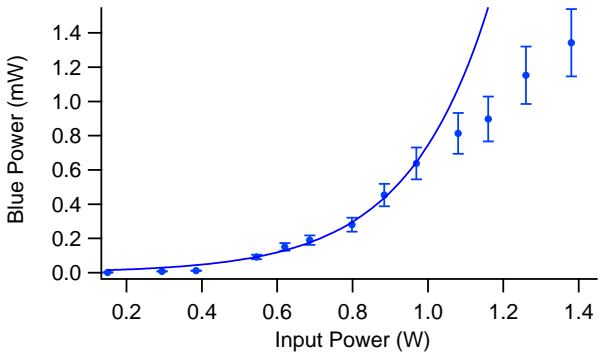}
\caption {The blue power generated as a function of the power in the excitation laser, with a Rb density of $3x10^{14} cm^{-3}$.  At low input powers, the expected exponential dependence is seen, with the dependence at higher densities trending toward linear.  The lower data points are fit to an exponential curve to help illustrate this change in dependence. }
\label{fig:intensity dependence}
\end{figure}

It is expected that the gain for the four-wave mixing process would go as the number of atoms in the $5d_{5/2}$ state, giving exponential dependence of the blue power on this population.  If the process is far from saturation, we also expect exponential dependence of the blue power on the excitation intensity.  We have measured the blue power generated as a function of the input power, with the data shown in  Fig. \ref{fig:intensity dependence}.  At low input powers, this dependence is consistent with exponential gain, but at high powers it trends toward a linear dependence.  This could be an indication that the process is approaching a saturation point, or that competing processes are becoming significant.  

\begin{figure}[htbp]
\centering
\includegraphics[width=8.6 cm]{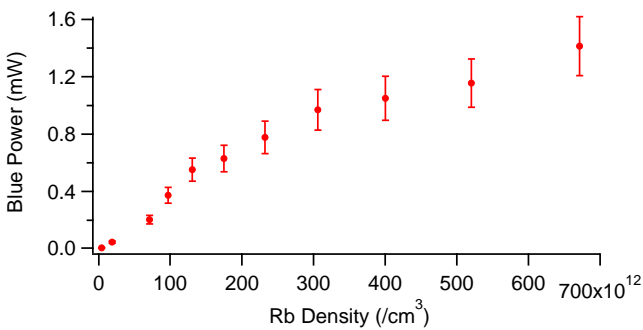}
\caption {The blue power generated as function of the rubidium density, with an excitation laser power of 1.4 W.  At high densities the blue power generated appears limited by competing processes or is approaching saturation.}
\label{fig:tempdependence}
\end{figure}

Changing the temperature of the cell allows control of the density of rubidium atoms, and allows further exploration of the possible onset of saturation.   Figure \ref{fig:tempdependence} shows the blue power as a function of the Rb density.  The density is limited, corresponding to temperatures under $180^{\circ}$C, in order to prevent alkali reaction with the cell walls. Here, the growth in blue power at high densities is clearly suppressed.  It has previously been observed that at higher temperatures competing processes may limit the successful production of blue light \cite{Brekke:13}. 

Both the intensity and temperature dependence suggest that we are approaching the saturation point of the four-wave mixing process.  Eventually, the process would be limited when the two-photon excitation rate through the 5p level, $\Omega^{(2)}_{5s\rightarrow5p\rightarrow5d}$, is equal to the two-photon excitation rate through the 6p level, $\Omega^{(2)}_{5s\rightarrow6p\rightarrow5d}$ \cite{Wunderlich:90}.  At this point the output power would only scale linearly with the input power.  Further investigation into the cause of saturation of the system remains an intriguing area of research.

Through the use of this build-up cavity, the 420 nm output power reaches as high as 1.9$\pm$0.3mW, comparable to the highest powers achieved with the two-step process \cite{Sell:14,Offer:16}.  This output power is 50 times the power achievable without the cavity.  Though there are signs of saturation, further gains could be made through increased input power, cavity reflectivity optimization, or using an additional cavity for the blue output \cite{Offer:16}.

\begin{figure}[htbp]
\centering
\includegraphics[width=8.8 cm]{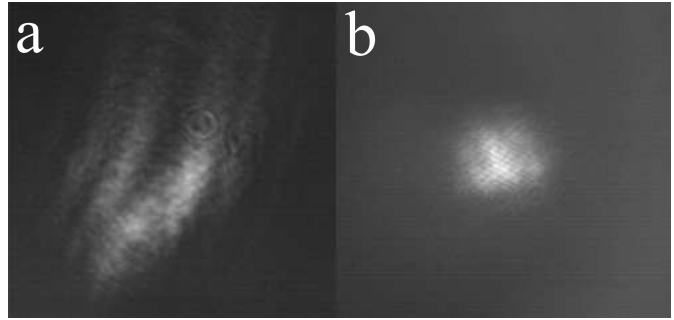}
\caption{a)Profile of the excitation beam after the tapered amplifier. b) Profile of the blue beam output from the ring cavity.}
\label{fig:profiles}
\end{figure}

In addition to achieving high output powers, the ring cavity is also beneficial in enhancing the spatial quality of the generated beams.  The tapered amplifier gives a non-gaussian beam with $M^2\approx2$, whose profile is shown in Fig. \ref{fig:profiles}a.

Without the use of the ring cavity, the generated 420 nm beam profile is also non-Gaussian.  As was pointed out in \cite{Sell:14}, blocking the edges of the input beam with an iris gives a $\approx10\%$ increase in the generated blue power.  A full explanation of this effect is still needed, but it may be that the outside portion of the beam generates regions where four-wave mixing cannot occur, but do generate counter-propagating, competing ASE.    

When the ring cavity is employed, the profile of the generated blue beam becomes an almost perfectly Gaussian beam, as shown in Fig. \ref{fig:profiles}b.  In addition, the iris no longer gives any appreciable gain in power, implying the cavity has already eliminated any outside portions of the beam not belonging to the Gaussian mode.  Thus, in addition to generating much higher powers, the output of the ring cavity is also of higher spatial quality for coupling into an optical fiber or for use in future experiments.

\section{Discussion and Conclusions}

The presence of the ring cavity provides significant power gains and an excellent spatial profile while maintaining a fairly simple experimental system.  There is still only one diode laser needed for the process, and the lack of temperature dependence to the phase matching criteria makes this system much easier to use than a standard frequency doubling crystal.  

The presence of the cavity does make it difficult to access the cell with an optical pumping beam, which has been shown to be successful elsewhere \cite{Akulshin:12b}.  This could still be accomplished either by slighlty non-collinear alignment of the pumping beam, or by using a cavity resonant for each frequency.  Even without these, however, the intensity output gained from the cavity is more than 10 times what could be expected from optical pumping.  

Recently it has been observed that a resonant cavity for the generated blue light increases output power and reduces the linewidth for the outgoing beam \cite{Offer:16}.  It could be a beneficial adaptation to construct a cavity which is resonant for both 778 nm and 420 nm around the rubidium cell.  The current work has not investigated the linewidth properties of the generated light, but this also remains an area of interest for the future.  

The output power from this system is comparable to that generated in the two-step scheme using lasers at 780 and 776 nm.  While the use of a cavity removes some of the simplicity of this format, it still makes use of a single excitation frequency and produces an easily controlled output frequency \cite{Brekke:15} due to the large intermediate state detuning.  If a sapphire cell is implemented, the resulting beam at 5.23 $\mu$m could be used, which is expected to have hundreds of $\mu$W available.  The resulting 420 nm beam is tunable around the rubidium $5s_{1/2}\rightarrow6p_{3/2}$ transition, and so presents an excellent candidate for use in future experiments where rubidium is excited to high principle quantum numbers.  

In conclusion, we have implemented a ring cavity to increase the circulating intensity of the two-photon excitation beam in rubidium.  The resulting increase in gain for parametric four-wave mixing generates $1.9\pm0.3$ mW of light at 420 nm, more than 50 times that possible without the cavity.  In addition, the generated beam is shown to have an excellent spatial profile.  The density and input power dependence of this process have been investigated, suggesting the process may be approaching saturation.

\bibliography{cavityfwm}


\end{document}